\documentclass[prl,aps,amsfonts,amssymb,amsmath,graphicx,showpacs,twocolumn,pdffig,amscd,ulem]{revtex4-1}

\usepackage{graphicx}

\usepackage{dcolumn}

\usepackage{bm}

\usepackage{natbib}

\usepackage{color}

\begin{document}


\title[]{On phase behavior and dynamical signatures of charged colloidal platelets}

\author{Sara Jabbari-Farouji $^{1,2}$, Jean-Jacques Weis $^{3}$, Patrick Davidson $^{4}$, Pierre Levitz $^{5}$, and Emmanuel Trizac $^{1}$}

\affiliation{$^{1}$
LPTMS, CNRS and Universit\'e Paris-Sud, UMR 8626, B\^at. 100,
91405 Orsay, France
}

\affiliation{$^{2}$
Laboratoire interdisciplinaire de Physique, UMR 5588, F-38041 Grenoble,
France
}

\affiliation{$^{3}$
LPT, CNRS and Universit\'{e} Paris-Sud, UMR 8627, B\^at 210, 91405 Orsay, France
}

\affiliation{$^{4}$
Laboratoire de Physique des Solides, CNRS and Universit\'{e} Paris-Sud, UMR 8502
B\^at 510, 91405 Orsay, France
}

\affiliation{$^{5}$
Laboratoire PECSA, UMR 7195, Universit\'e Pierre et Marie Curie,
 Case Courrier 51, 4 place Jussieu, 72522 Paris Cedex 5, France
}

\date{\today}


\begin{abstract}
 We investigate the competition between anisotropic excluded-volume and repulsive electrostatic interactions in suspensions of thin charged colloidal discs,
 by means of Monte-Carlo simulations and dynamical characterization of the structures found. We show that the original intrinsic anisotropy of the electrostatic potential
 between charged platelets, obtained within the non-linear Poisson-Boltzmann formalism, not only rationalizes the generic features of the complex phase diagram of 
 charged colloidal platelets such as  Gibbsite and Beidellite clays, but also predicts the existence of novel structures. In addition,  we find evidences of a strong slowing down
 of the dynamics upon  increasing density.
\end{abstract}

\maketitle

 Recent breakthroughs in the synthesis of particles with precisely controlled shapes 
and interactions have sparked a renewed interest in understanding the influence of particle anisotropy on phase behavior
\cite{Glotzer,Glotzer1}, which is required to achieve well-controlled self-assembly and design of novel structures.
In the simplest case of axially symmetric particles, anisotropic excluded-volume interactions lead to formation of
liquid-crystalline phases \cite{Onsager} while more complex shapes result in remarkable structural diversity
\cite{Glotzer1}. Directional interactions between particles, induced by their geometry
and chemical structure, can also lead to anisotropic complex self-assembled structures.

 When both particle geometry and interactions are anisotropic, the interplay, and even competition between the excluded-volume 
and other (e.g. electrostatic, van der Waals) interactions may lead to the emergence of novel structures and geometrical frustration. Charged platelet 
suspensions, such as clays \cite{Laponite,bentonite,Beidellite,nontronite},
disc-like mineral crystallites \cite{Gibphase, zirconium} or exfoliated nanosheets \cite{nanosheets,nanosheets1}, that have been
 widely studied experimentally provide such
examples and are the subject matter of this report. All these systems are highly charged discotic colloids.  Some of them, like Laponite and Bentonite
 clay suspensions \cite{Laponite,bentonite}, form arrested states with increasingly slow dynamics at low densities. Others, like Beidellite and Gibbsite
\cite{Beidellite,Gibphase}, exhibit an equilibrium isotropic-nematic transition at moderate densities. 

These observations raise fundamental 
questions about the influence of interparticle interactions on the isotropic-nematic transition
and more generally on the organization of charged platelets. Here, we aim at disentangling the generic effects due to shape and charge, from more specific ones 
such as attractive interactions \cite{Tanaka,Ruzicka} that depend on details of chemical constitution of
each platelet system. We consequently focus on the role of  repulsive electrostatic interactions on the phase behavior, 
although the attractive interactions turn out to
be important for formation of gel-like arrested states in some systems like Laponite clays at low densities \cite{Ruzicka,SaraPRL}.

For hard-core repulsions only, upon increasing platelet density, a sequence of isotropic, nematic, columnar, and crystalline
phases appears, depending on the aspect ratio \cite{cutspheres,Marechal}. Despite several simulation works
\cite{Dijkstra, Kutter,Mossa,Labbez,Wensink}, the influence of electrostatic interactions on phase behavior is still poorly
understood and a comprehensive study that provides an overview and rationalization of the experimental data is missing.
Previous studies suffer from not taking into account the renormalised effective charge and the full orientational dependence of interactions \cite{Dijkstra,Wensink}.
Furthermore, most of them are based on coarse-grained interaction site models \cite{Kutter,Mossa,Labbez}, which are computationally costly.

 Here, such shortcomings are circumvented: our approach is based on a well controlled effective electrostatic potential between two discs
 obtained within Poisson-Boltzmann formalism \cite{Trizac,Carlos}. 
 We address two important questions: i) how do repulsive electrostatic interactions, in conjunction with hard core, influence the isotropic-nematic (I/N) transition?
 ii) how do these interactions affect the translational and rotational dynamics of charged discs upon varying density or ionic strength, and 
 can they lead to a slow dynamics?
\begin{figure}[h]
\includegraphics[scale=0.2]{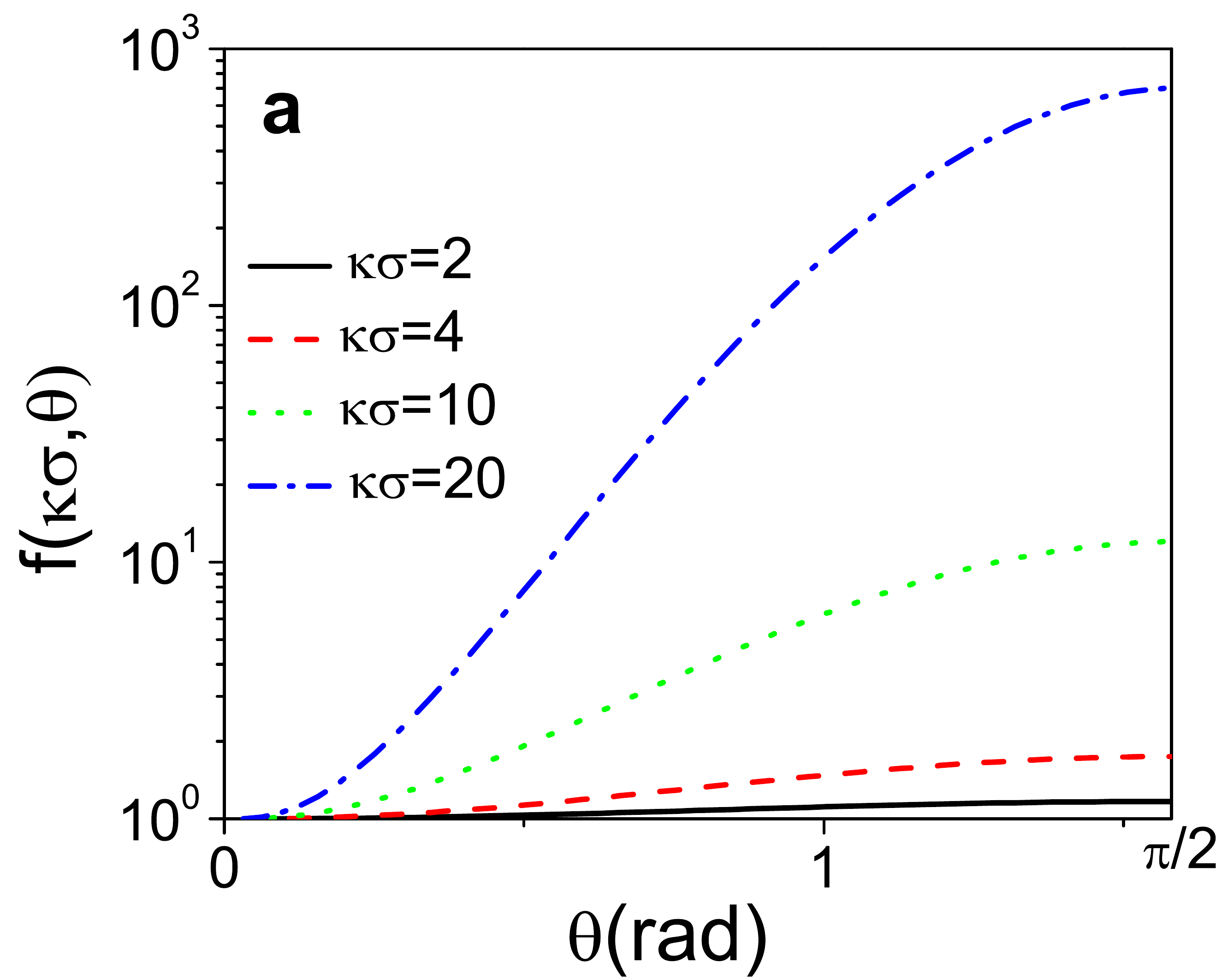}
\includegraphics[scale=0.2]{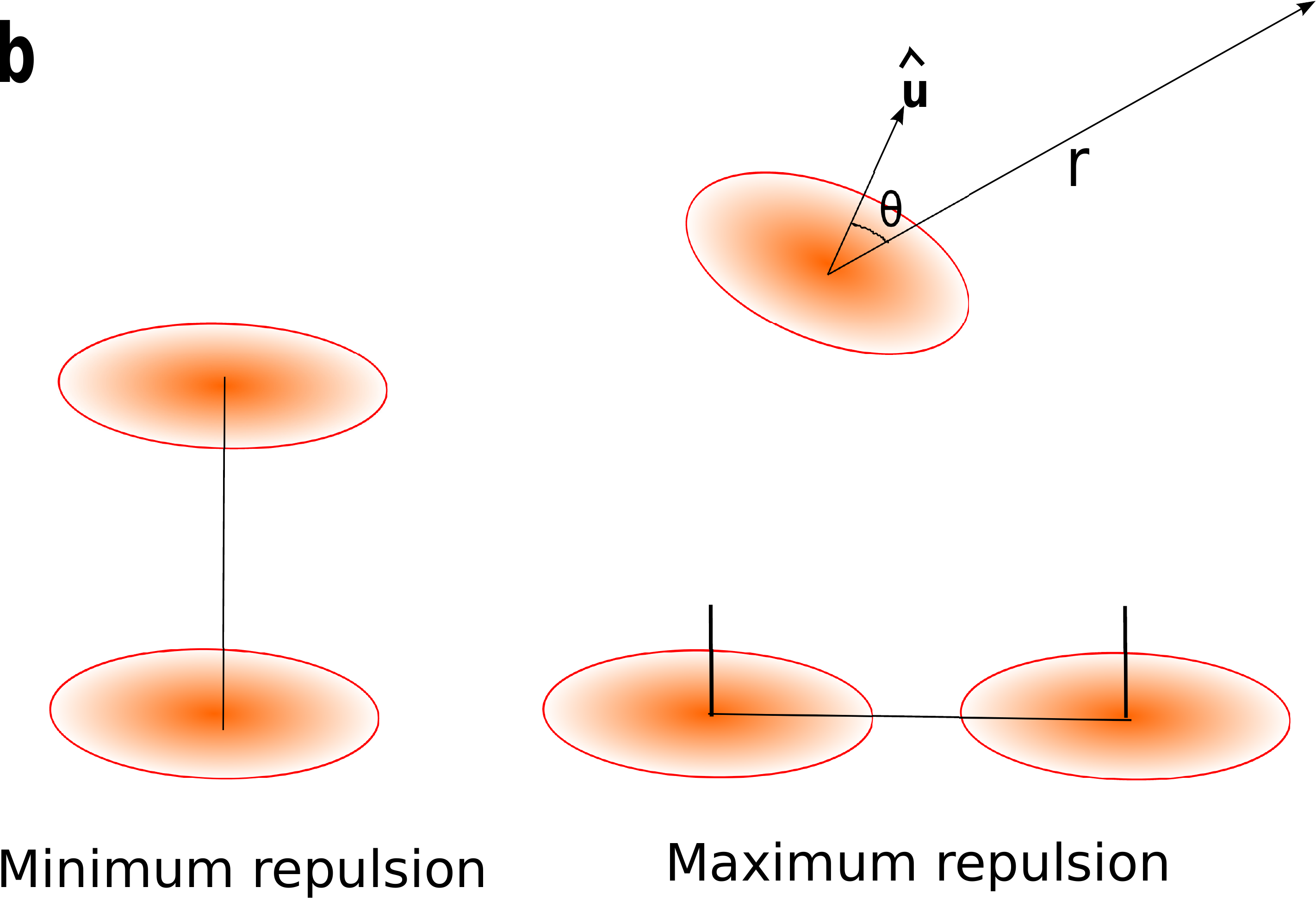}
\caption{ a) Anisotropy function $f(\kappa \sigma, \theta)$ versus $\theta$, the angle between the disc normal $\widehat u$ and the line 
of observation, for several values of  $\kappa \sigma$ b) Two interacting parallel charged discs in stacked and coplanar 
configurations with equal center to center distance correspond to minimum and maximum repulsion, respectively. }
\label{figZ}
\end{figure}

The far-field dimensionless electrostatic interaction potential between
two charged discs of diameter $\sigma$ immersed in an electrolyte medium with ionic strength  $I=\kappa ^2/ (8 \pi \lambda_B)$,  centers at a distance $r$ apart, and 
with normals making angles $\theta_i$ with the center-to-center line, reads:
\begin{equation} \label{eq:pot2}
 U_{12}(r/\sigma,\theta_1, \theta_2) = \lambda_B Z_{eff}^{2} (\kappa \sigma) f(\kappa \sigma, \theta_1) f(\kappa \sigma, \theta_2) e^{- \kappa r}/r.
\end{equation}
Here, $\kappa$ is the inverse Debye length, $\lambda_B$ is the Bjerrum length (7.1 \AA\ in water at room
temperature) and $Z_{eff}$ is the effective charge; it depends on the platelet bare charge and  salt through $\kappa \sigma$.
For highly charged platelets as considered here, $Z_{eff}$ saturates to a value $Z_{eff}^{sat}\lambda_B/\sigma=0.5\kappa \sigma+1.12$. 
The anisotropy function  $f(\kappa \sigma, \theta)$ encodes the orientational dependence of the potential
and can be expressed in terms of spheroidal wave functions \cite{Carlos}. $f(\kappa \sigma, \theta)$ increases with angle $\theta$, reaching a
 maximum in the disc plane ($\theta=\pi/2$) (Fig. \ref{figZ}a). It  is furthermore enhanced by the ionic strength ($\propto \kappa^2$): a smaller Debye length ($1/\kappa$) acts 
as a probe which reveals asphericity. This results in a strong asymmetry between coplanar and stacked configurations (Fig. \ref{figZ}b),
a notable difference with standard Heisenberg-like models. A competition between the excluded volume and electrostatic effects
ensues, which is a generic feature for charged oblate spheroids.  The interplay between anisotropy in shape and interactions 
has important consequences on the phase behaviour and dynamics.
\begin{figure*}[ht]
\includegraphics[scale=0.55]{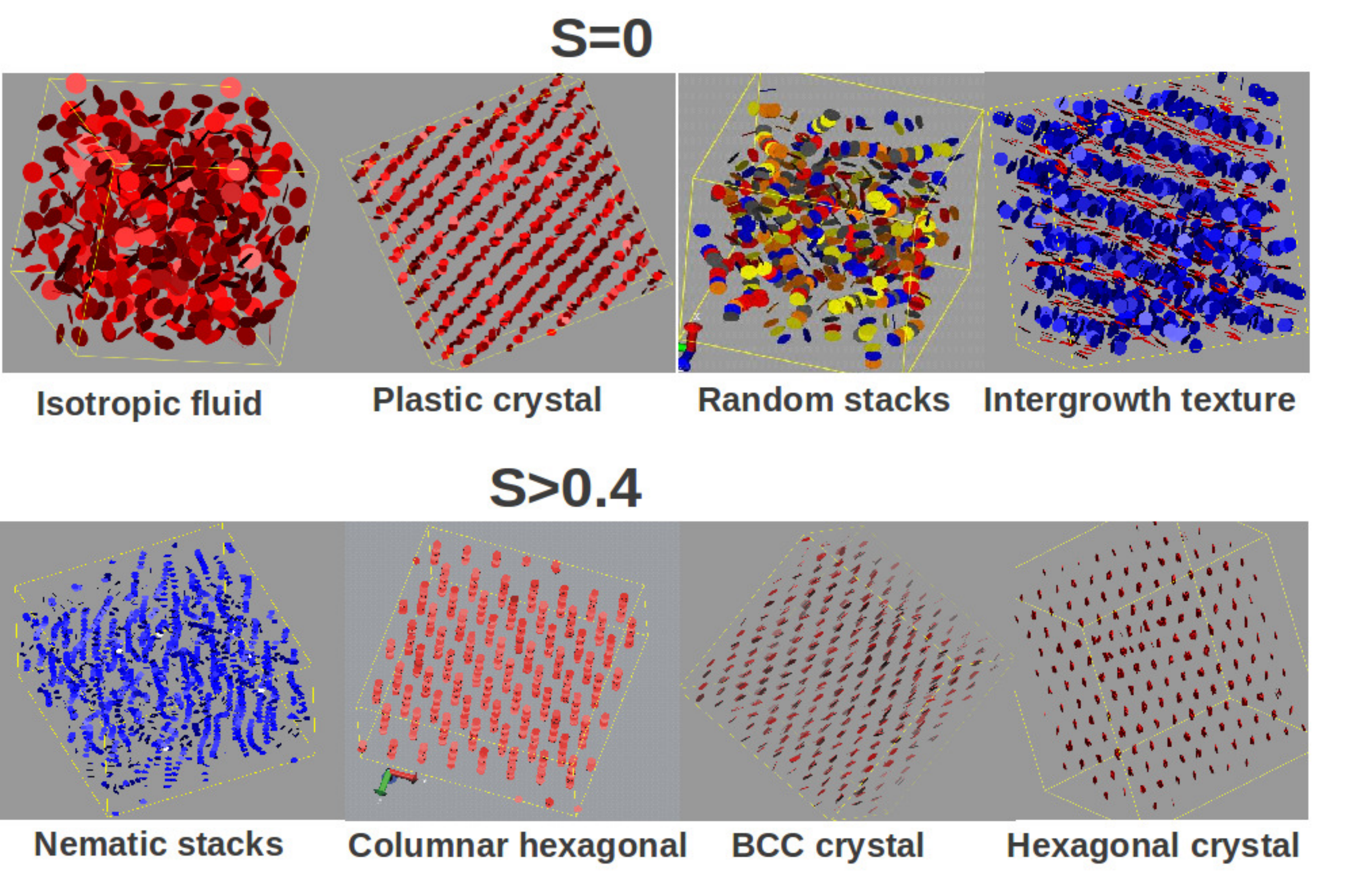}
\caption{We recognize eight distinct configurations for charged discs, four of which have a vanishing orientational order parameter $S$. These include isotropic fluid, plastic crystal, "random stacks" and a novel "intergrowth texture" that is composed of discs arranged in alternating nematic
(red colour) and antinematic (blue color) layers. The orientationally ordered structures with $S> 0.4$ encompass nematic
 and columnar hexagonal liquid-crystals in addition to two crystalline phases with bcc and hexagonal symmetry. Note that
the discs are not displayed with their real size, but with 20 to 40 \% of their actual diameter.}
\label{struct}
\end{figure*}

We start by providing some details on the simulation techniques.
We have investigated the structure of hard discs interacting by the
anisotropic Yukawa potential of Eq. (\ref{eq:pot2}) by Metropolis Monte-Carlo simulations \cite{Allen} which generate states according to appropriate Boltzmann weights.
 We carried out canonical ($NVT$) simulations on a 
system of $N$ thin hard discs of
diameter $\sigma$ in a cubic simulation box of side $L$ with periodic boundary conditions using the anisotropic
two-body potential of Eq. (\ref{eq:pot2}). The simulations were performed for a wide range of densities $\rho^*=N \sigma^3/L^3=0.1-8$ and
 ionic strengths giving rise  to  $\kappa \sigma=1-20$.
The dimensionless diameter $\sigma/\lambda_B$ was taken to be 43 in the simulations, although we shall argue that its precise value, in a parameter 
range to be specified, is irrelevant. The number of particles varied between $10^3 \leq N \leq 10^4 $, although most of simulations were performed with $N=1372$ discs corresponding to 
box sizes ranging from $L/ \sigma=17.0$ for $\rho^*=0.1$ to $L/ \sigma=5.5$ for $\rho^*=8$.
For  $\kappa \sigma \leq 2 $, we adopted an Ewald-like scheme (neglect
of the small Fourier space contribution) to take into account the relatively
long range of the potential \cite{Ewald}. For  $\kappa \sigma=2$ in the density range $\rho^*= 1-4 $, we confirmed that Ewald-like simulations performed with  $ N=1024 $ particles
lead to the same results as  with larger box sizes ($N=4000-10976$, $ L/\sigma \approx 15 $) without Ewald sums.

Due to relatively large values of the effective charge leading to a
complex free energy landscape, we sometimes encountered, at large
densities, $\rho^* \geq 3$, dependence of the final configuration on
initial conditions.
We circumvented this problem by  simulated annealing \cite{SimuAnneal} i.e. we started from an initial equilibrium configuration of (uncharged) 
discs at the desired density, then increased the charge gradually from
zero to the final value $Z_{eff}^{sat} (\kappa \sigma) $.
 We  performed (depending on the density and $\kappa \sigma$)
$0.3-1 \times 10^6$ MC cycles, (a cycle consisting of translation and
rotation of the $N$ particles) at each equilibration step before increasing the charge value. 
This gradual charge increment, similar to gradual cooling of the
system for a fixed value of the effective charge, leads to reproducible
results for independent simulation runs performed with different initial
conditions, provided the
charge increment between two subsequent stpdf is small enough, typically
$\Delta Z_{eff} \lambda_B / \sigma \leq 0.5 $. 



It is worthwhile here to emphasize that a precise determination of the phase diagram by performing free energy calculations, would be difficult as 
 the true free energy of the system, in addition to the free energy of  the  colloids, would have
to include the contribution stemming from the colloid-microions interactions \cite{free1,free2}  which cannot be  simply obtained from
the effective colloid-colloid pair potential. For this reason, we identified the different phase regions in density-charge
space from their structural behavior obtained in the $NVT$ calculations.

We further performed dynamic Monte-Carlo simulations \cite{DMCsara} to obtain the intermediate scattering function and time orientational 
correlation functions of orientationally disordered systems. In this case, very small values of displacements were used and we performed $0.1-2 \times 10^7$ MC cycles to obtain  the dynamical quantities.  The degree of orientational order was characterized by the nematic order parameter
$ S\equiv \left\langle P_2(\cos (\psi)\right\rangle$ where the second Legendre
polynomial reads $P_2(x)=\left(3 \cos^2 x -1) \right/2$, $\psi$ is the angle of a platelet normal with
the director $\widehat{n}$ and the brackets mean averaging over all particles. Hence, $-1/2<S<1$ with $S = 0$
for the disordered isotropic phase. When $0 < S < 1$, the platelet normals point
on average along the director $\widehat{n}$ while in the more unconventional $S <0$ regime, the platelet normals are on average perpendicular
to $\widehat{n}$ ("antinematic" order).

Upon varying density and ionic strength (proportional to $(\kappa \sigma) ^2$), we found eight distinct structures (Fig. \ref{struct}). Four of them have a vanishing $S$ and are
formed at low and moderate densities. Three of these structures, isotropic fluid, plastic crystal (structure with positional
order but no orientational order), and "random stacks", are orientationally disordered while the fourth one is a novel structure
that consists of intertwined nematic and anti-nematic layers (Fig. \ref{struct}). We 
coin it "nematic-antinematic intergrowth texture". The orientationally ordered phases with $S> 0.4$ which appear at higher densities
include two liquid-crystalline phases, i.e. nematic phase of platelet stacks (sometimes called "columnar nematic") and hexagonal
 columnar phase, together with two crystalline
phases with bcc-like and hexagonal symmetry.
\begin{figure}[h]
\includegraphics[scale=0.3]{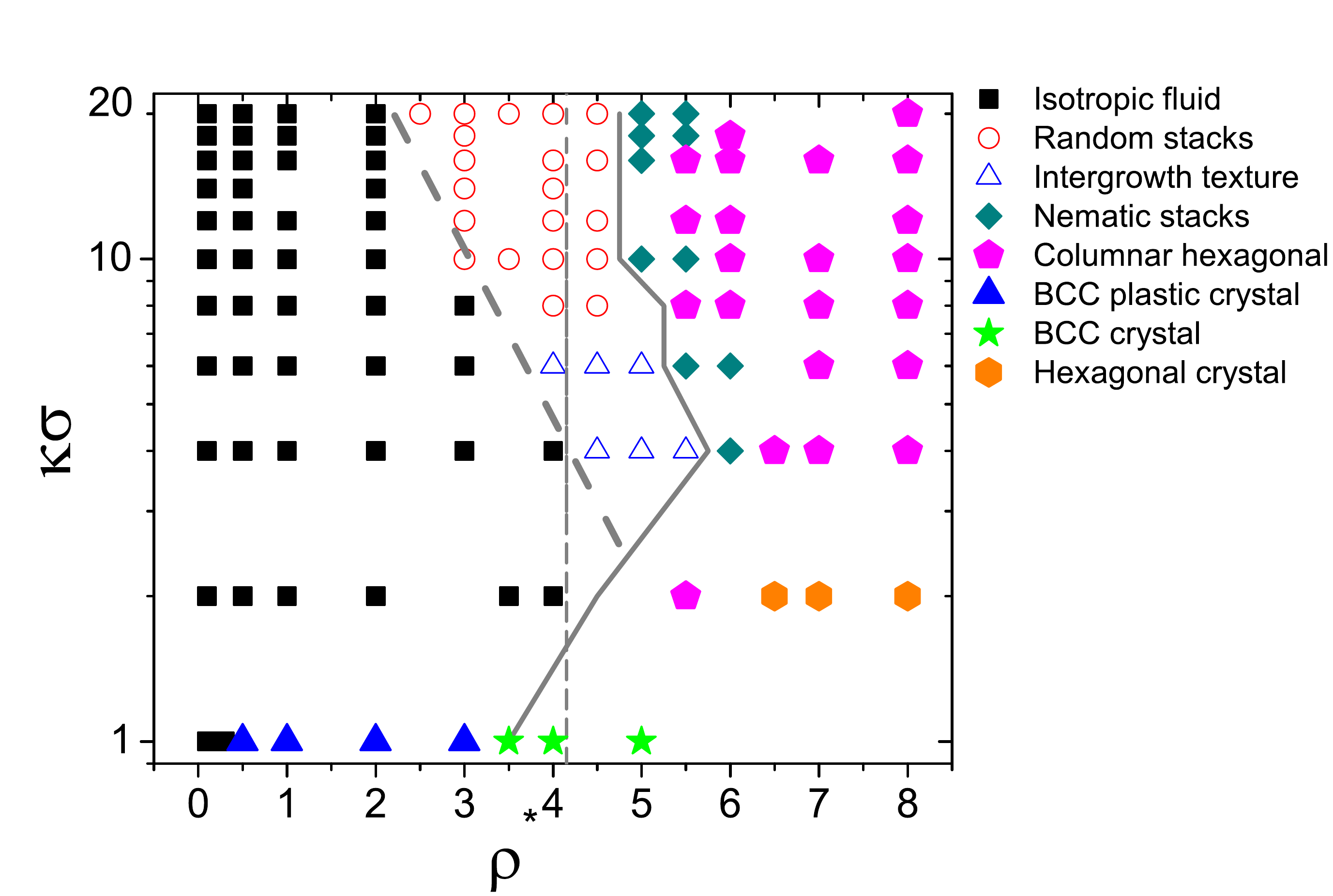}
\caption{ Phase diagram of repulsive charged discs as a function of dimensionless density $\rho^*$ and  $\kappa \sigma$ which is proportional to square root 
of ionic strength. The vertical dashed line at $\rho^*=4.2$ shows the density beyond which the nematic phase
for hard discs (thus without electrostatic interactions) appears. The solid line separates the orientationally ordered
(right hand side) and disordered phases (left hand side). The oblique dashed line is a guide to the eyes to
 show, on the right hand side, the regions where the structures with slow orientational dynamics appear.}
\label{phase}
\end{figure}

Before turning to a comparison with experiments, we present the phase diagram. We emphasize that
our gradual equilibration procedure for $\rho^*=\rho \sigma^3 \geq 3$ allows us to attain configurations close to equilibrium.
More rapid quenches generally put the system out of equilibrium, and for high enough densities, we therefore
expect discrepancies with the experimentally observed states, that have presumably not reached equilibrium.

We now highlight important aspects of the phase diagram (Fig. \ref{phase}) in connection
with the features of our pair potential. A significant hallmark is the non-monotonic
behavior of orientational disorder-order transition with $\kappa \sigma$ which results from the opposing effects of the
decreasing potential range and the increasing amplitude of the anisotropy function, upon increasing ionic strength; 
$Z_{eff}^{sat}$ also increases  but we verified
that the observed trend is unchanged provided $Z_{eff}\lambda_B/\sigma >2.3$. This in turn implies that
 the particular value of $\sigma/\lambda_B$, which is {\it a priori} a relevant dimensionless parameter,
in practice plays little role. This opens the possibility to discuss the results pertaining to different particle sizes
 solely in terms of $\rho^*$ and $\kappa \sigma$, as will be done below.

 In the limit of low ionic strengths corresponding to $\kappa \sigma=1 $, where the potential is rather isotropic and long-ranged,
bcc-like structures are formed, as observed with charged spheres
at low $\kappa \sigma$ \cite{Dijkstra1}. At low densities, discs are orientationally disordered (plastic crystal)
and are reminiscent of the Wigner crystals
 observed for low volume fractions of charged spheres \cite{chaikin}. Increasing the ionic strength, at
 $\kappa \sigma=2$ where the amplitude of the anisotropy function is less than 1.2, the plastic crystal is replaced by an isotropic
 fluid at low and moderate densities. The crystal disappearance for a slightly larger
 $\kappa \sigma$ and shorter range of potential highlights the geometrical frustration effect of anisotropy.
At $\kappa \sigma=2$, in the high density regime, the long-range positional order is still preserved as crystals with hexagonal
symmetry appear. Interestingly, for large enough ionic strengths leading to $4 \leq \kappa \sigma \leq 6$ and at moderate densities, the
new intergrowth texture appears and, for $\kappa \sigma \geq 8$, randomly oriented stacks of discs are observed. The intergrowth
texture consists of sets of aligned discs ($0.75 < S < 0.93$) interspersed with layers exhibiting anti-nematic
order ($-0.45 < S < -0.3$). Both types of layers share the same director.
Here, the particles organize with relative T-shape configurations, intermediate between coplanar and stacked.
At still higher densities, further increase of ionic strength equivalent to $\kappa \sigma \geq 4$, leads to weakening of the positional
order. Hexagonal crystals are replaced by hexagonal columnar liquid-crystals with no positional correlations along the nematic director.
\begin{figure}[h]
\includegraphics[scale=0.32]{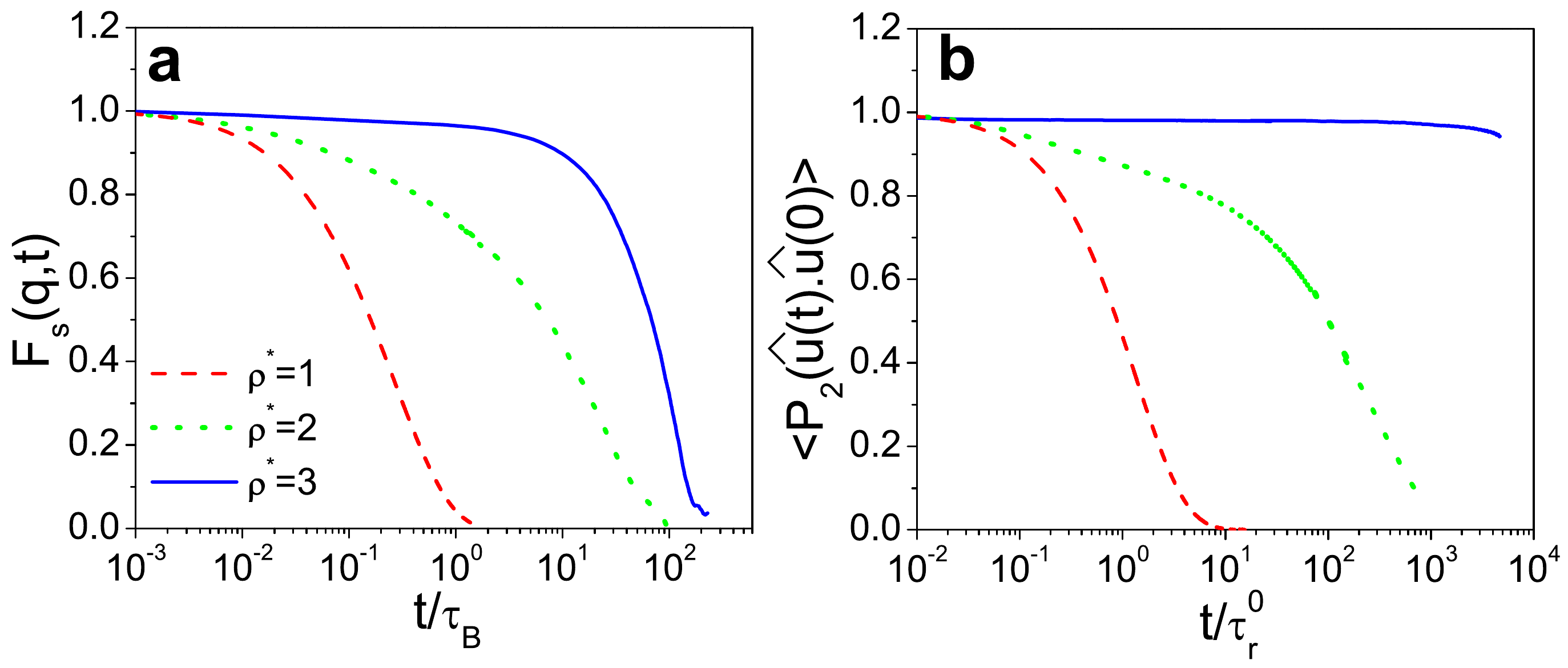}
\caption{ a) Self-intermediate scattering function
$F_s(q,t)$ computed at $q\sigma=7$ and b) orientational time correlation function $\left\langle P_2(\widehat{u}(t)\cdot \widehat{u}(0)) \right\rangle$
for   densities $\rho^* =1- 3$  at $\kappa \sigma=20$. The time-scales
are expressed in terms of characteristic time of free translational $ \tau_B \equiv \sigma^2/ (6 D^t_{0})$ and
rotational $\tau_{0}^{r}\equiv1/(6 D_{0}^r) $ diffusion coefficients, respectively.}
\label{dynamics}
\end{figure}

\begin{figure}[h]
\includegraphics[scale=0.2]{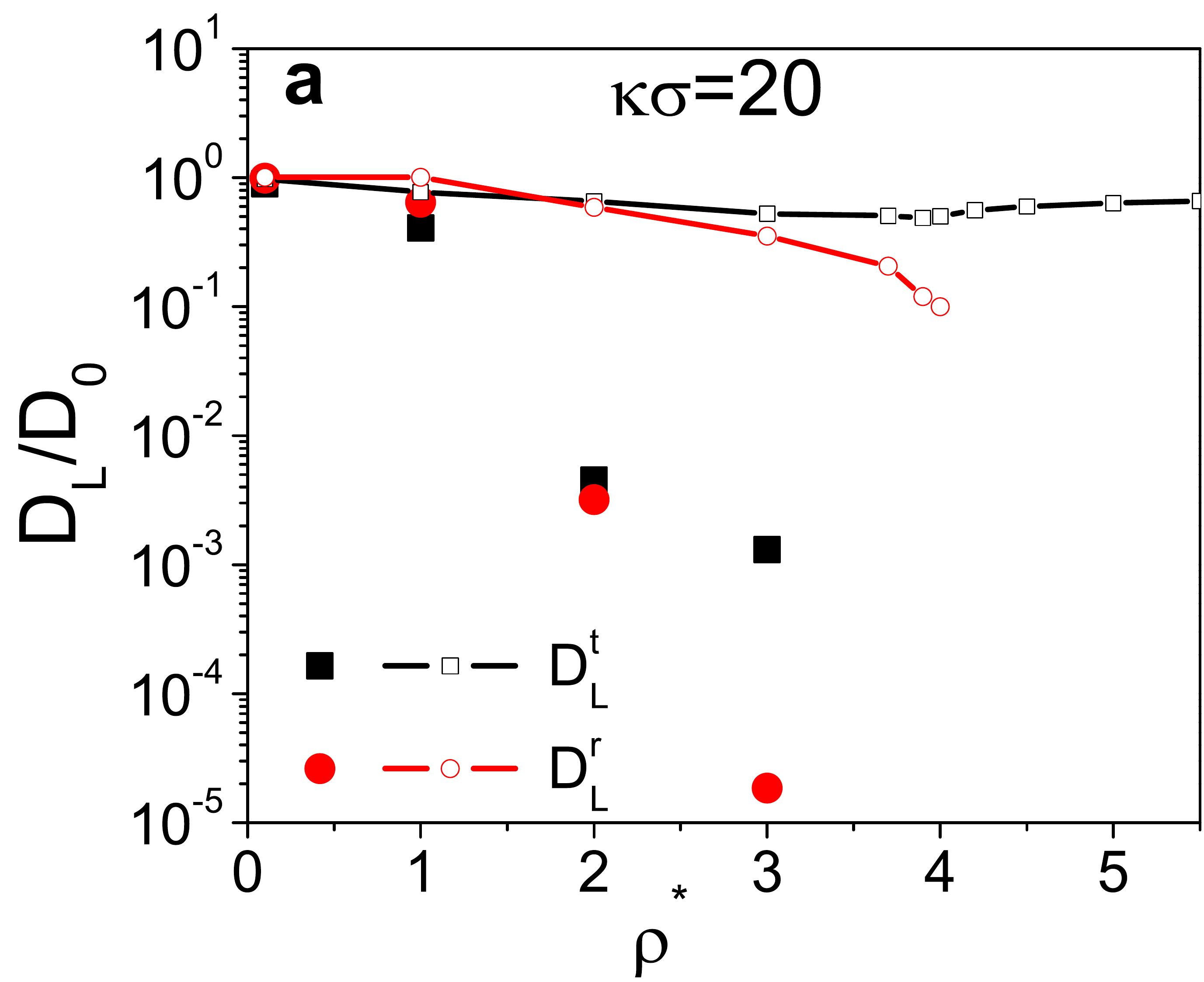}
\includegraphics[scale=0.2]{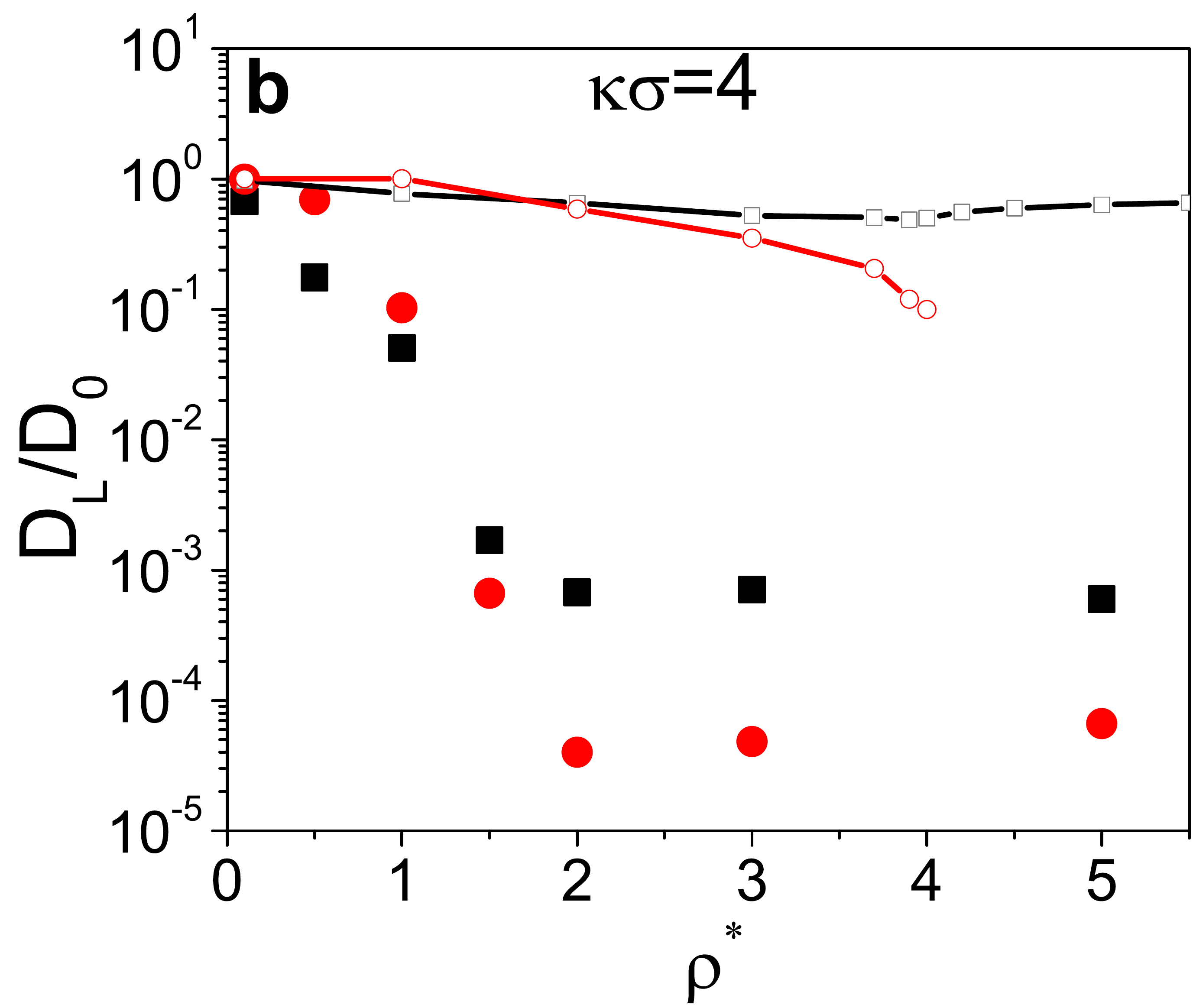}
\caption{ The long time translational diffusion coefficient $D_L^t$ and rotational diffusion coefficient  $D_L^r$   of charged discs normalized by their value in 
the infinite dilution limit,
as a function of density at $\kappa\sigma=20$ (panel a, filled symbols) and  $\kappa\sigma=4$ (panel b, filled symbols). For comparison, we
have also shown the long-time diffusion coefficients of hard discs (empty symbols). }
\label{diffusion}
\end{figure}

To better characterize the orientationally disordered structures, we examined their translational and rotational dynamics by the dynamic
Monte-Carlo method \cite{DMCsara}.  In Fig. \ref{dynamics}, we have shown the  self-intermediate scattering functions
$F_s(q,t)=\left\langle \exp[i \vec{q}\cdot (\vec{r}(t)-\vec{r}(0)) ]\right\rangle$ and time orientational correlation functions
$\left\langle P_2(\widehat{u}(t)\cdot \widehat{u}(0))\right\rangle$  computed  for densities $\rho^*=1-3$, lower than those of 
the nematic phase at $\kappa \sigma=20$. Upon increasing the density, both translational and rotational relaxation times increase. Notably, 
 in the random stacks phase, the orientational time correlation functions exhibit very little decay. This shows that upon approaching the nematic phase, the 
 orientational relaxation  becomes dramatically slow (Fig. \ref{dynamics}b). In order to quantify the slowing down of dynamics, 
 we show in Fig. \ref{diffusion} the long-time translational and rotational diffusion coefficients, extracted from self-intermediate scattering function (equivalently from mean-squared displacement) and orientational
 correlation function \cite{DMCsara} at $\kappa \sigma=4$  and 20. We find that increasing the density, both long-time translational and 
 rotational diffusion coefficients of charged discs (filled symbols)  strongly decrease, with much  steeper slope than the corresponding ones for hard discs (empty symbols). 
 Also, in the limit of very low ionic strength, the plastic crystals exhibit slow dynamics for translational coordinates and a fast decay for orientational ones (not shown here).

 At  higher densities where random stacks and intergrowth texture form, the spectacular slowing down of dynamics upon increase of density as a result of  repulsive electrostatic interactions must have important consequences on 
 viscoelastic properties of such charged  platelet suspensions. One expects this strong slowing down of dynamics  to be concomitant with a huge increase of viscosity, 
 as found in  dynamical measurements \cite{Gibbdynamics,Sarasoftmatter},  see the discussion below.

We now compare our simulations against the most extensively studied
charged colloidal platelets: Gibbsite, and Beidellite systems for which the repulsive interactions are predominant.
Gibbsite and Beidellite show an I/N transition
for a wide range of ionic strengths, and platelet stacking is often observed in the nematic phase \cite{Gibphase,Beidellite}. Moreover, Gibbsite suspensions
also display a columnar hexagonal liquid-crystalline phase. These features are well captured in our simulations.

\begin{figure}[h]
\includegraphics[scale=0.3]{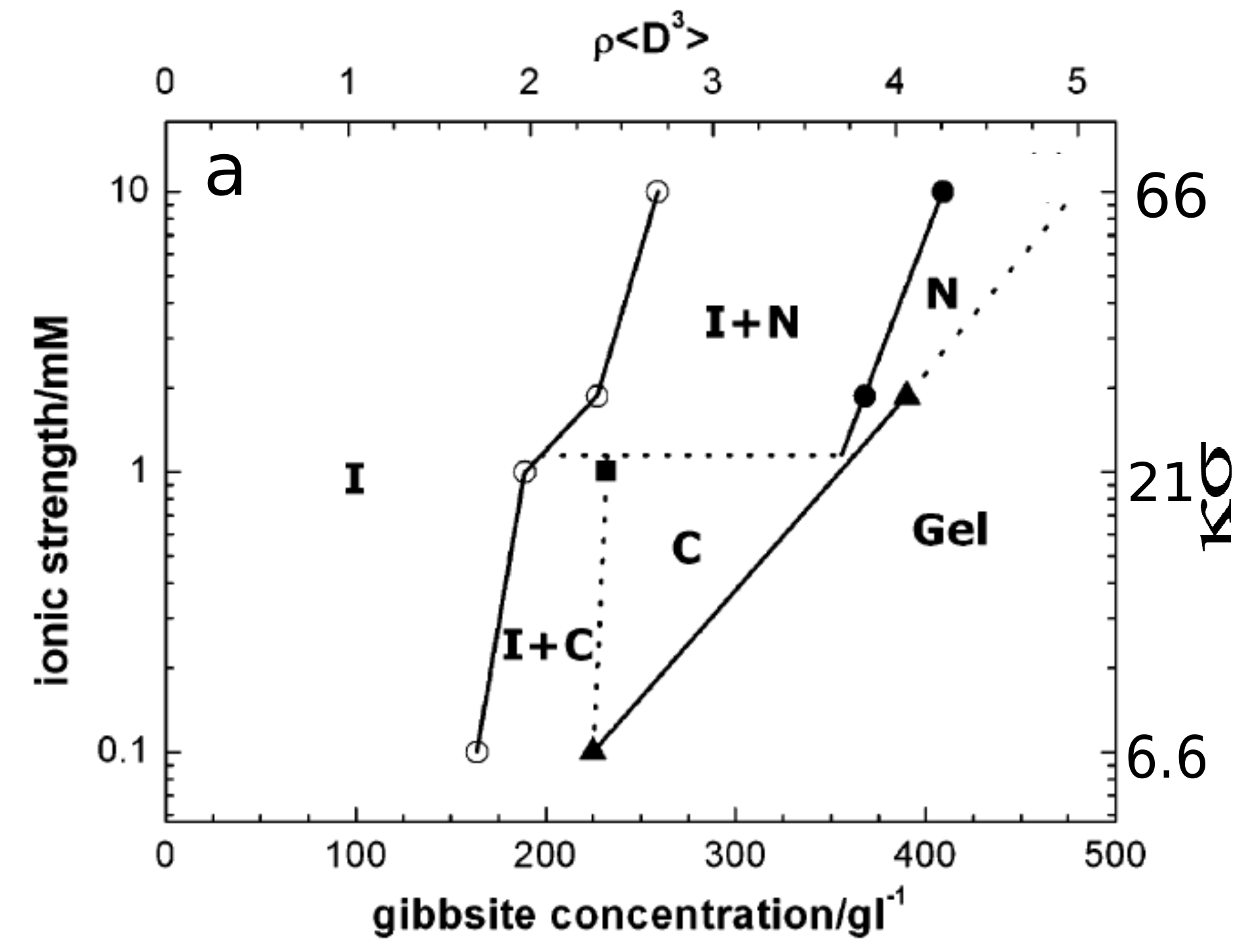}
\includegraphics[scale=0.2]{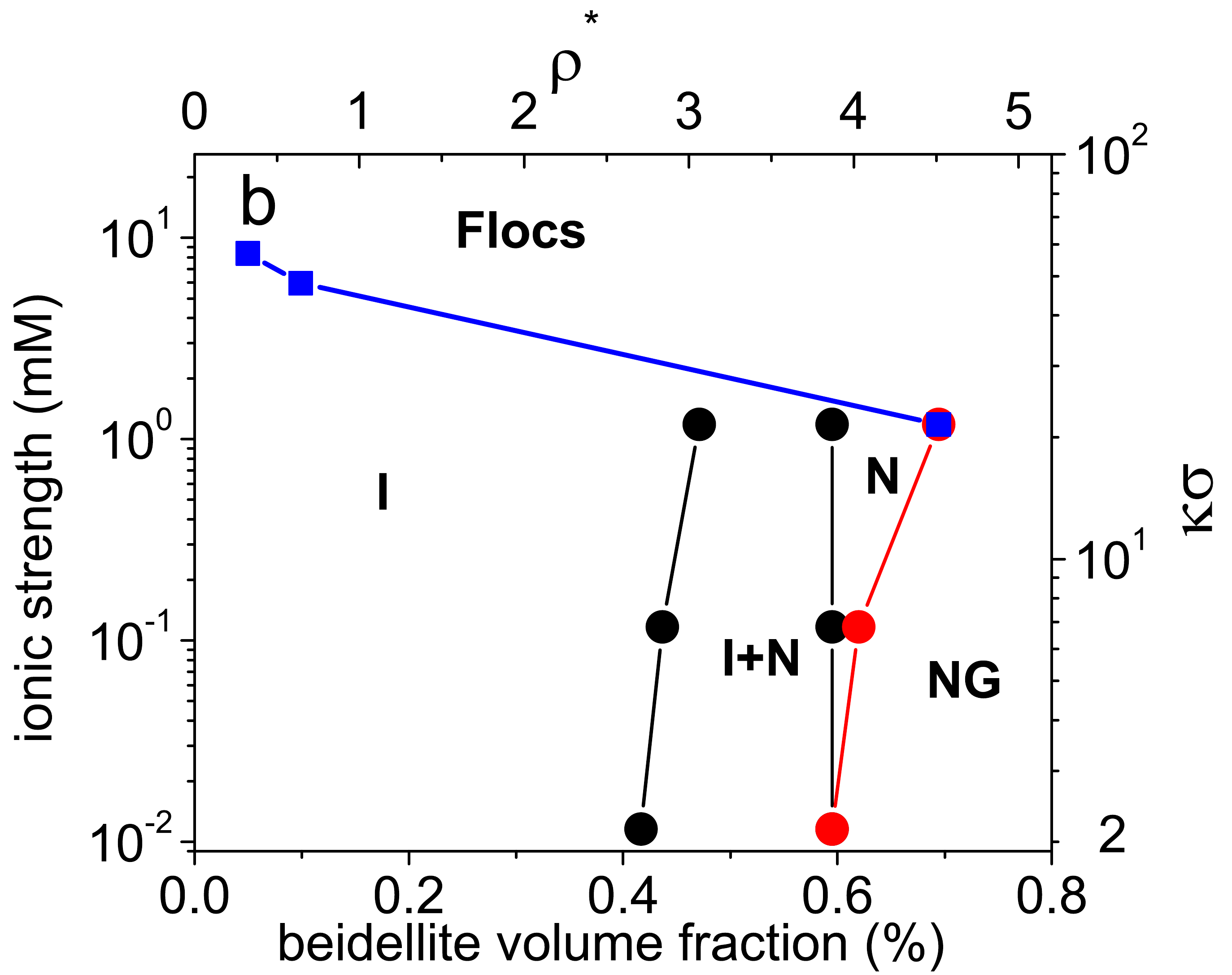}
\caption{ Experimental phase diagrams of a) Gibbsite from  (Fig 4 in ref.  \cite{Gibphase1}) and b) Beidellite from 
(Fig. 7 (c) in \cite{Beidellite}). The phases corresponding to each symbol are indicated in the legend. Here, C  refers to Columnar, I to  isotropic,
N to Nematic (stacks) phase and NG to nematic Gel.}
\label{experiments}
\end{figure}

Fig. \ref{experiments} shows the phase diagrams of Gibbsite \cite{Gibphase1} and Beidellite \cite{Beidellite} versus concentration/volume fraction and ionic strength 
 (bottom/left axes) accompanied by the corresponding reduced density $\rho^*$ and $\kappa \sigma$ (top/right axes) for quantitative comparison. Overall, the agreement between 
 our simulations and the experimental phase diagrams for Gibbsite \cite{Gibphase} and Beidellite \cite{Beidellite} (shown in Fig. \ref{experiments})
is good. In both Gibbsite and Beidellite, for $10 <\kappa \sigma <20$, a nematic phase of stacked platelets is found,
although at densities   lower than the one obtained in simulations. This  discrepancy could be due to platelet polydispersity that
usually widens the isotropic-nematic coexistence region. Moreover, in agreement with our simulations, 
 a direct transition from isotropic to columnar hexagonal phases is observed for Gibbsite particles at  lower ionic strengths.
Upon further increase of density, a nematic gel is experimentally observed whereas we obtained a columnar hexagonal phase.
At such high densities, we already anticipated that the suspensions cannot easily reach equilibrium, since the
simulations require a slow and careful procedure to relax. Furthermore, the nematic gel of
Gibbsite is a jammed state of oriented stacks (columns) that evolves towards a columnar hexagonal structure with time \cite{gib2008}.
In addition, the recent measurements of long-time translational and rotational diffusion of isotropic suspensions of Gibbsite in DMSO \cite{Gibbdynamics} are qualitatively in agreement with
the results shown in  Fig. \ref{diffusion}.  Although the ionic strengths for these experiments are not reported, we observe  a similar trend for both $D_L^t$ and $D_L^r$ that decrease by more than 3 orders of magnitude upon approaching the isotropic-nematic transition.

At this point, it is worth discussing our simulation results in connection with Laponite clay, another system of highly charged platelets. 
In contrast to Gibbsite and Beidellite, the I/N transition in Laponite suspensions is hindered by the occurrence of various
kinetically trapped states at very low volume fractions ($\phi <0.02$). The range of densities and ionic strengths investigated in experiments corresponds
to $\rho^* < 1$ and $1 \leq \kappa \sigma <20$. At such low densities, we observe a solid-like state (plastic crystal) only at small
$\kappa \sigma=1$. This state could presumably only be glassy if polydispersity were considered \cite{Zaccarelli}. Upon further increase
of ionic strength, we observe an ergodic repulsive isotropic liquid in the low-density region of our phase diagram. 
Interestingly, we find that onset of slow dynamics shifts towards lower densities   upon increase of ionic strength, similar to the trend observed for Laponite.
Our model system only based on repulsive interactions cannot fully explain the phase behavior of Laponite for larger ionic strengths;
these discrepancies point to the relevance of attractions \cite{Ruzicka} or other specific interactions in this system. Nevertheless, our approach
hints at the existence of a solid-like state at low ionic strengths, in agreement with the experimental observation of a Wigner
glass \cite{Levitz,Wigner}.

In conclusion, numerical simulations of charged platelets, interacting with an effective potential obtained from
Poisson-Boltzmann theory, show a rich phase diagram that captures the generic features of the experimental systems
and  provides us an insight about the influence of electrostatics on slowing down of dynamics upon varying density and ionic strength. The strong decrease of long time translational and rotational diffusion found in our simulations is similar to the trend observed for  Gibbsite suspensions \cite{Gibbdynamics}. 
The required procedure of slow charge increase, at moderate and high densities, echoes the difficulty for
experimental systems to reach thermodynamic equilibrium. Better agreement with experiments could arguably be achieved by
considering polydispersity, platelet flexibility, van der Waals attractions, and interactions specific to each system such as solvation forces.
The features brought to the fore here, though, can be considered as the non-specific effects pertaining to charged colloidal
 platelets. In particular, the robust form of our screened Coulomb potential
leads to the formation of novel intergrowth texture, which calls for further experimental 
investigations in the isotropic region of the phase diagram, close to the I/N transition.\\

\begin{acknowledgments}
\noindent\small{S. J-F would like to acknowledge the financial support from Foundation Triangle de la Physique and IEF Marie-Curie fellowship.
We are also grateful to A. Maggs and  H. H. Wensink for fruitful discussions.}
\end{acknowledgments}


\begin{thebibliography}{99}



\bibitem{Glotzer}  Glotzer, S. C. and  Solomon, M. J.,  \textit{Nature materials} \textbf{2007} 6, 557--562.

\bibitem{Glotzer1} Damasceno,  P. F.; Engel,  M. and  Glotzer, S. C.,   \textit{Science} \textbf{2012} 337, 453--457.

\bibitem{Onsager}  Onsager, L. , \textit{Ann. N. Y. Acad. Sci.} 51 \textbf{1949},  627--659.



\bibitem{Laponite} Ruzicka, B. and  Zaccarelli, E. , \textit{Soft Matter} \textbf{2011} 7, 1268--1286.

\bibitem{bentonite}  Gabriel, J.-C. P.;  Sanchez, C. and  Davidson, P., \textit{J. Phys. Chem.} \textbf{1996} 100, 11139--11143 .

\bibitem{Beidellite} Paineau, E. et al. \textit{J. Phys. Chem. B} \textbf{2009}, 113, 15858--15869.

\bibitem{nontronite} Michot,  L.J.; Bihannic, I.;  Maddi, S.;  Baravian,  C.;  Levitz, P. and  Davidson,  P.,   Langmuir \textbf{2008}, 24, 3127 



\bibitem{Gibphase}  Mourad, M. C. D.; Byelov  D. V.; Petukhov, A. V.; Matthijs de Winter,  D. A.;  Verkleij, A. J. and  Lekkerkerker, H. N. W.

   \textit{J. Phys. Chem. B} \textbf{2009} 113, 11604--11613.

\bibitem{zirconium}  Sun, D.;  Sue, H.-J.;   Cheng, Z.,  Martínez-Ratón,  Y. and Velasco, E.,  \textit{Phys. Rev. E} \textbf{2009}, 80, 041704.

\bibitem{nanosheets}  Liu, Z.,  et al.,   \textit{J. Am. Chem. Soc.} \textbf{2006}, 128 (14), 4872--4880. 

\bibitem{nanosheets1}  Yamaguchi, D., et al. \textit{Phys. Rev. E} \textbf{2012}, 85, 011403.



\bibitem{Tanaka}

Tanaka,  H.;   Meunier, J.  and  Bonn, D., \textit{Phys. Rev. E} \textbf{2004} 69, 031404.

\bibitem{Ruzicka}

Ruzicka, B.; Zaccarelli, E.;  Zulian, L.;  Angelini, R.;  Sztucki, M.;  Moussaid, A.;  Narayanan, T.  and Sciortino, F.,

\textit{Nature Materials} \textbf{2011}, 10, 56 .

\bibitem{SaraPRL}Jabbari-Farouji, S.;   Wegdam, G. H. and Bonn,  D., Phys.  \textit{Rev. Lett.} \textbf{2007} 99, 065701.

\bibitem{cutspheres} Veerman, J. A. C. and  Frenkel, D. Phase behavior of disklike hard-core mesogens. \textit{Phys. Rev. A} \textbf{1992}, 45, 5632--5648.


\bibitem{Marechal} Marechal,  M.;   Cuetos, A.;  Martinez-Haya,  B.,  and  Dijkstra, M.

 \textit{J. Chem. Phys.} \textbf{2011}, 134, 094501.




\bibitem{Dijkstra} Dijkstra,  M.; Hansen, J. P.  and  Madden, P. A., \textit{Phys. Rev. Lett.} \textbf{1995}, 75, 2236--2239.

\bibitem{Kutter}  Kutter, S.;  Hansen, J.; Sprik,  M. and  Boek, E.,  \textit{J. Chem. Phys.} \textbf{2000}, 112, 311--323.

\bibitem{Mossa}  Mossa, S.;  de Michele, C.;  Sciortino,  F.,  \textit{J. Chem. Phys.} \textbf{2007}, 126, 014905.

\bibitem{Labbez}Delhorme,  M.;  J$\ddot{o}$nsson, B. and  Labbez, C., \textit{Soft Matter} \textbf{2012}, 8, 9691--9704.

\bibitem{Wensink} Morales-Anda, L.; Wensink, H. H.;   Galindo, A. and  Gil-Villegas, A.

  \textit{J. Chem. Phys. 136} \textbf{2012}, 034901.

\bibitem{Trizac}  Agra, R.; Trizac, E. and   Bocquet, L., \textit{Eur. Phys. J. E} \textbf{2004}, 15, 345--357.

\bibitem{Carlos}  \'{A}lvarez, C. and T\'{e}llez,  G.,   \textit{J. Chem. Phys.} \textbf{2010}, 133, 144908.

\bibitem{Allen} M. P. Allen and D. J. Tildesley, \textit{Computer Simulation of Liquids} (Oxford University Press, Oxford, 1987).

\bibitem{Ewald} Salin G. and  Caillol, J. M., \textit{J. Chem. Phys.} \textbf{2000},  113, 10459--10463.

\bibitem{SimuAnneal}  Kirkpatrick, S.;   Gelatt, C. D. and  Vecchi, M. P. \textit{Science} \textbf{1983}, 220, 671--680.

\bibitem{free1} Dobnikar, J.;  Castañeda-Priego, R.; von Grunberg,  H. H.;  and   Trizac, E., \textit{New Journal of Physics} \textbf{2006}, 8, 277.

\bibitem{free2}  Trizac, E.;  Belloni, L.;  Dobnikar, J.;   von Grunberg; H. H., and Castaneda-Priego,  R. \textrm{Phys. Rev. E} \textbf{2007}, 75, 011401.

\bibitem{DMCsara}  Jabbari-Farouji, S. and  Trizac, E.,   \textit{J. Chem. Phys.} \textbf{2012}, 137, 054107.

\bibitem{Dijkstra1}  Hynninen, A-P. and Dijkstra, M.,   \textit{Phys. Rev. E} \textbf{2003}, 68, 021407.

\bibitem{chaikin}  Lindsay, H. M. and Chaikin,  P. M.,   \textit{J. Chem. Phys.}  \textbf{1982}, 76, 3774--3782.




\bibitem{Gibphase1}  van der Beek, D. and  Lekkerkerker, H. N. W.,  \textit{Langmuir} \textbf{2004}, 20, 8582--8586.



\bibitem{gib2008} Mourad, M. C. D.; Byelov,  D. V.;  Petukhov,  A. V. and  Lekkerkerker, H. N. W.,

\textit{J. Phys. Condens. Matter} \textbf{2008} 20, 494201.



\bibitem{Gibbdynamics}  Kleshchanok, D.; Heinen, M.;  Nagele, G. and  Holmqvist, P. , \textit{Soft Matter}, \textbf{2012}  8, 1584--1592.

\bibitem{Sarasoftmatter}  Jabbari-Farouji,S.;  Zargar, R;.  Wegdam, G; and  Bonn, D. ; \textit{Soft Matter}, \textbf{2012}, 8, 5507--5512.

\bibitem{Zaccarelli} Zaccarelli,  E.;  Andreev,  S.;   Sciortino, F. and Reichman,  D. R.,

 \textit{Phys. Rev. Lett.} \textbf{2008}, 100, 195701.







\bibitem{Levitz} Levitz, P.;  Lecolier, E.;  Mourchid, A.;  Delville,  A. and Lyonnard, S., 

\textit{Europhys. Lett.} \textbf{2000}, 49, 672--677.



\bibitem{Wigner}  Bonn, D.;  Tanaka,  H.;  Wegdam, G.;  Kellay, H. and  Meunier, J. \textit{Europhys. Lett.} \textbf{1998}, 45, 52--57.





\end{thebibliography}
\end{document}